\documentclass[a4paper,preprint,superscriptaddress,preprintnumbers,nofootinbib]{revtex4}
\usepackage{amsmath,amssymb}
\usepackage{bm}
\usepackage{graphicx}

\newcommand{\Be}{\begin{eqnarray}}
\newcommand{\Ee}{\end{eqnarray}}
\begin{document}
\title{Restricted Boltzmann Machines for the Long Range Ising Models}
\author{Ken-Ichi \surname{Aoki}}
\email{aoki@hep.s.kanazawa-u.ac.jp}
\affiliation{Institute for Theoretical Physics, Kanazawa University, Kanazawa 920-1192, Japan}
\author{Tamao \surname{Kobayashi}}
\email{kobayasi@yonago-k.ac.jp}
\affiliation{National Institute of Technology, Yonago College, Yonago 683-8502, Japan}
\preprint{KANAZAWA-16-04}
\begin{abstract}
We set up Restricted Boltzmann Machines (RBM) to reproduce the Long Range
Ising (LRI) models of the Ohmic type in one dimension.
The RBM parameters are tuned by using the standard machine
learning procedure with an additional method of 
Configuration with Probability (CwP).
The quality of resultant RBM are evaluated through the susceptibility with respect to 
the magnetic external field.
We compare the results with those by Block Decimation Renormalization Group
(BDRG) method, and our RBM clear the test with satisfactory precision.

\end{abstract}
\maketitle

\section{Introduction}
Quite recently the deep learning machines have drawn much attention
since they are very effective for image processing \cite{Xie12} and Go game \cite{Go16} etc.
The deep learning machines can be regarded as a method of
information reduction keeping as much macroscopically 
important features as possible.
This policy or idea resembles to the renormalization group method
in physics \cite{Wilson75,NPRG}, and the intrinsic relationship between them 
has been argued \cite{VRG-DL}, and further investigation is strongly desired from
both sides.

Here in this article we make Restricted Boltzmann Machines (RBM) 
\cite{Smo86,Hinton02} to reproduce
the Long Range Ising (LRI) models in one dimension. 
The LRI models have its own 
long history \cite{LRI-old}, since they work as most simplified models to investigate
the quantum dissipation issues which are still unveiled subjects lying in between
classical and quantum physics 
\cite{Caldeira-Leggett,suzuki92,MC-qm,Instanton-qm, nprg-qm} . 
The LRI models have a critical 
point (temperature) to exhibit the spontaneous magnetization \cite{math-qm,KT} . 
The critical
point depends on the long range nature determined by the power exponent
of the interactions \cite{MC-Ising} .
Moreover, functional renormalization group approaches revealed
Ising model critical phenomena\cite{Wetterich,Delamotte}.
They organized new framework using the field-theoretic 
formulation and developed powerful techniques with 
including higher order diagrams. In this article, however, 
we are attempting to utilize the finite range scaling method
and direct comparison of our results with the above
mentioned ones are not possible yet.
Here we adopt the Ohmic case where the power exponent 
is $-2$ and it is known to give a marginal point of bringing about 
the finite critical temperature ($K_{1\rm c}=0.657$) \cite{Ising-FRS08} .

We are interested in the procedure of how we can make up appropriate
RBM to reproduce the LRI models.
We define RBM with link parameters respecting the translational
and Parity invariance.
We generate sample data set of the LRI models, and make 
RBM to learn the data by tuning the RBM parameters to reach the
maximum likelihood point.

Here we introduce a new method in setting up the input data for
RBM learning.
We slice out the learning procedures into many small steps, where
the sample data set is defined by a set of pair of Configuration
with its corresponding Probability (CwP).
This method resembles to the multi-canonical ensemble method
where a part of the Hamiltonian is moved out of the configuration
measure into the physical quantity to be averaged.

Finally we evaluate the total quality of the tuned up RBM.
Starting with completely random spins, 
we produce output set of configurations, 
step by step, checking the susceptibility of each set.
After the onset of equilibrium, we examine the susceptibility. 
We compare our results with those calculated
by a renormalization group method
called the Block Decimation Renormalization Group (BDRG) which gives
the exact susceptibility numerically \cite{Ising-FRS08,QM-FRS12}.
This article is a short report of our work and full analysis will be
published elsewhere.
 
\section{Restricted Boltzmann Machines for Long Range Ising Models}

We introduce the standard RBM consisting of 
visible variables $\bm{v}$ and hidden variables $\bm{h}$.
The total probability distribution is defined by 
\Be
P(\bm{v},\bm{h})=\frac{1}{Z} e^{-H(\bm{v},\bm{h})}\ ,\ \ 
Z=\sum_{\bm v},\bm{h}  e^{-H(\bm{v}, \bm{h})}\ ,
\Ee
where $H$ is the Hamiltonian (energy function)
 and the partition function $Z$ is the total 
normalization constant.
We integrate out the hidden variables to get the probability 
distribution function for $\bm{v}$,
\Be
P(\bm{v})=\sum_{\bm{h}} P(\bm{v},\bm{h})\ .
\Ee
The standard restriction of RBM requires the Hamiltonian to take the
following form,
\Be
H(\bm{v},\bm{h})=-\sum_{i, j} w_{i j} v_i h_j\ ,
\Ee
where we omit the external field terms (linear in $\bm{v},\bm{h}$) here.

\begin{figure}[htbp]
    \centering
    \includegraphics[clip, width=10cm]{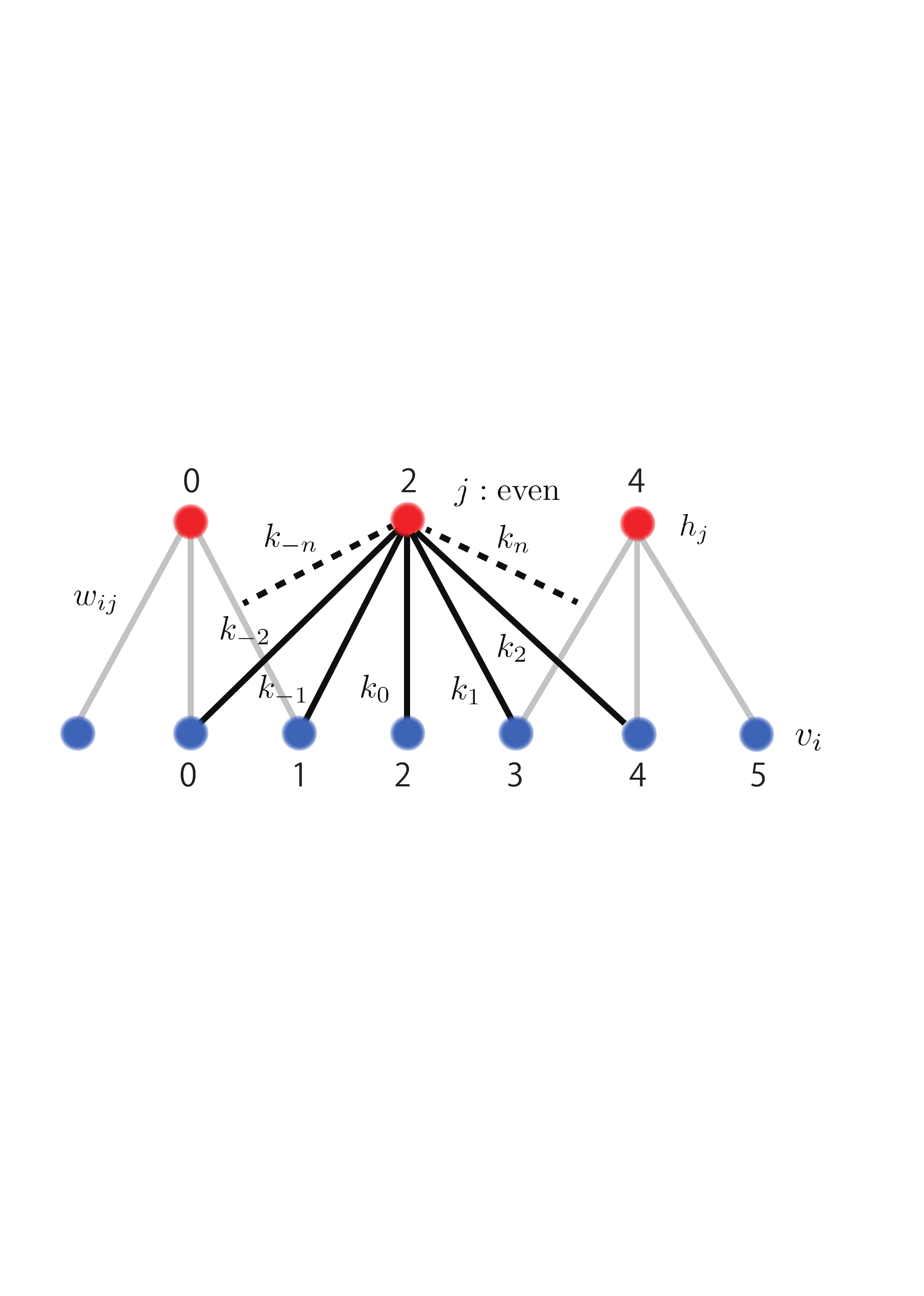}
    \caption{Definition of Restricted Boltzmann Machines.}       
    \label{fig:1layer-RBM}
\end{figure}

Our target system is one dimensional Ising system and
all variables $\bm{v},\bm{h}$ takes $+1$ or $-1$ respectively.
The number of hidden variables is exactly half of that of visible
variables and we adopt the periodic boundary condition.
The links and their weights are defined as drawn in Fig.~1.
Although other types of linking pattern may be considered,
we take this type since it has a well-defined nearest neighbor
solution. 
We respect the translational and Parity invariance of the system, 
and therefore the link weights are all common to each hidden variables
and also are left-right symmetric $k_{-n}=k_n$.
Note that precisely speaking, the translational invariance holds for the
hidden sector only, and in the visible sector, odd and even site spins
are not equivalent.
Hereafter our RBM  are denoted by
\Be
P(\bm{v},\bm{h};\bm{k})\ ,
\Ee
where the machine parameter $\bm{k}$ represents $\{k_0, k_1, \cdots\}$.
The RBM visible probability distribution is given similarly,
\Be
P(\bm{v};\bm{k})=\sum_{\bm{h}} P(\bm{v},\bm{h};\bm{k})\ .
\Ee

\section{The Long Range Ising Models}

The LRI model is defined by the following 
statistical weights,
\Be
P_{\rm LRI} (\bm{v})=\frac{1}{Z}\exp\left(\sum_n K_n\sum_i v_i v_{i+n}
\right)  ,
\Ee
where $K_n$ is the coupling constant for range $n$.
We take the Ohmic type of the long range behavior,
\Be
K_n = \frac{K_1}{n^2} \ .
\Ee
Our purpose in this article is to tune the RBM parameter $\bm{k}$ 
so that the RBM visible probability
distribution may best reproduce the LRI probability distribution, 
\Be
P(\bm{v};\bm{k^*}) \simeq P_{\rm LRI}(\bm{v})\ .
\Ee

We divide the LRI Hamiltonian into two parts which are the nearest neighbor
base part and other long range part,
\Be
P_{\rm LRI}  &=& \frac{1}{Z} \exp\left( K_1\sum_i v_i v_{i+1}\right) 
\exp\left(\sum_{n=2} K_n\sum_i v_i v_{i+n}\right)\nonumber\\
&=& \frac{1}{Z} \exp(-H_0(\bm{v})) \exp(-H_{\rm L}(\bm{v}))\ .
\Ee
We set up the input data for RBM as follows.
In this section we omit the Hamiltonian slicing procedure for simplicity, 
which will be explained in the next section.
We generate a set of spin configurations exactly respecting the
nearest neighbor part probability distribution. 
This is done most quickly via the domain wall representation \cite{DWRG09} 
where the domain wall exists with probability,
\Be
q=\frac{1}{1+\exp(2K_1)}\ ,
\Ee
and there is no correlation among domain walls. Therefore we can
set each domain wall independently, expect for caring the periodic boundary condition.
We express this set of configuration by $\{\bm{v}^{(\mu)}\}$ 
where $\mu$ denotes discriminator of each configuration.

Then the second part of the weight is considered as the
physical quantity side. We calculate the additional {\sl probability} which should
be assigned to each configuration generated above,
\Be
\bm{v}_\mu \Longrightarrow p_\mu \propto \exp(-H_{L}(\bm{v}_\mu))\ .
\Ee
Now our target probability distribution to be learned by RBM 
is defined by a set of pair of
Configuration with corresponding Probability, CwP:
\Be
\{ \bm{v}^{(\mu)}; p_\mu\}\ .
\Ee
The normalization of the probability is taken to be 
\Be
\sum_\mu p_\mu = N\ ,
\Ee
where $N$ is the total number of configurations.

Now we define the likelihood of RBM to produce the above 
CwP as follows:
\Be
L(\bm{k}) =\prod_{\mu}^{N} p_\mu\sum_{\bm{h}}P({\bm v}^{(\mu)} ,\bm{h};\bm{k}).
\Ee
Note that the probability $p_\mu$ is included representing the effective number 
of occurrences of the corresponding configuration.
To search for the stationary point of the likelihood, we differentiate the
logarithm of likelihood function with respect to $\bm{k}$.
Using the explicit definition of our RBM, we have the following
derivative,
\Be
\frac{1}{Nj_{\rm M}}\frac{\partial \log(L(\bm{k}))}{\partial k_{n}}
&=&\frac{1}{Nj_{\rm M}}\sum_{\mu}^{N} p_\mu\sum_j^{j_{\rm M}}
v_{j+n}^{(\mu)} \tanh (\lambda_j)
-\frac{1}{j_{\rm M}}\sum_j^{j_{\rm M}}E\left[ v_{j+n} h_j ;\bm{k} \right]\ ,
\Ee
where $j_{\rm M}$ is the total number of hidden variables $\bm{h}$,
$\lambda_j$ is defined by
\Be
\lambda_j= \sum_n k_n v_{j+n}^{(\mu)}\ ,
\Ee
and $E[\ \cdot\ ]$ denotes the expectation value of operator
by the RBM,
\Be
E\left[ v_{j+n} h_j ;\bm{k} \right]
= \sum_{\bm{v},\bm{h}}  v_{j+n} h_j P(\bm{v},\bm{h};\bm{k})\ .
\Ee

Using the derivative above we adopt the steepest descent method
to find the maximum likelihood position of RBM parameters.
The expectation value part is evaluated by the contrastive divergence
method with several times of sample updates \cite{Hinton02}.

\section{Machine Learning Procedure and Results}

Our purpose here is to make appropriate RBM to generate high quality
distribution of spin chain for the 1D Ising model with long
range interactions. If the interactions among spins are
limited to the nearest neighbor type, the model is easily
solved exactly and the corresponding RBM solution 
is also obtained straightforwardly, although the practical machine
learning process is not trivial. However if the interactions are not
nearest neighbor, the model cannot be solved analytically, and
its RBM counterpart is far from trivial.

\begin{figure}[htbp]
    \includegraphics[clip, width=15cm]{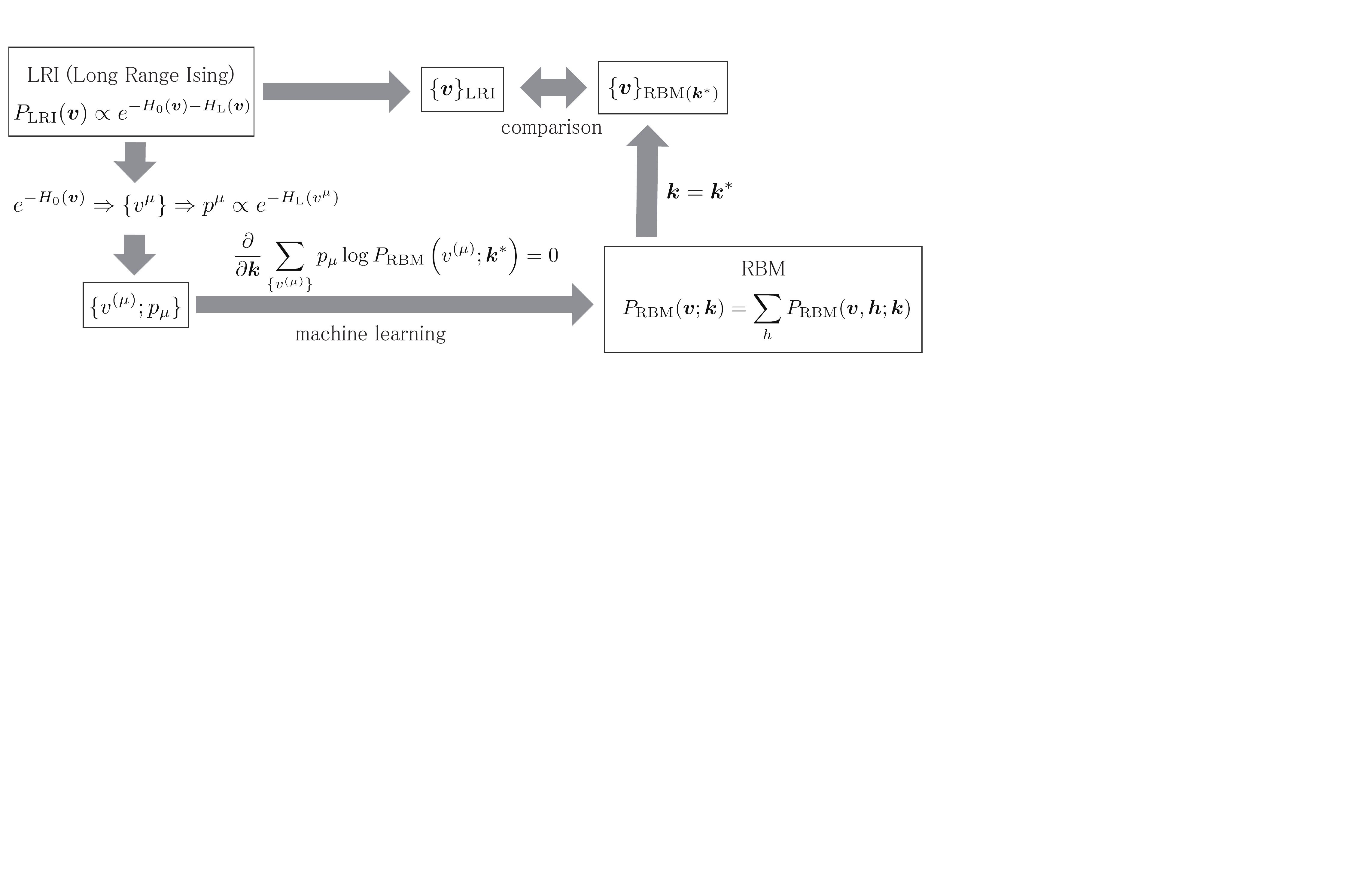}
    \caption{View of RBM learning procedures.}       
    \label{fig:image}
\end{figure}

Our total strategy here is drawn in Fig.~2, although we omit
the slicing processes explained below for simplicity. Slicing is necessary since
the difference of the probabilities in a set should not be too large.
The large deviation of probabilities causes 
drastic loss of sample quality and the effective size of data set is 
shrunk. We tune the slicing width 
so that the averaged probability might be limited by some small size.

Now the slicing procedure is explained in some detail.
First of all we prepare sliced Hamiltonians $\Delta H_m \ (m=1,2,\cdots m_{\rm Max})$
so that they satisfy the following properties:
\Be
\sum_{m=1}^{M} \Delta H_m (\bm{v}) = H_M(\bm{v})\ ,\ \ 
\left. H_M(\bm{v}) \right|_{M=m_{\rm Max}}=H_{\rm L} (\bm{v}).
\Ee
At each slicing step ($m$-th step here), the initial input configurations, denoted by
\Be
\{\bm{v}^{(\mu)}[m]\},
\Ee
is regarded as sarisfyng the probability distribution,
\Be
P_m(\bm{v}) \propto \exp(-H_0(\bm{v})) \exp(-H_{m-1}(\bm{v})).
\Ee
Then we assign additional probability factor given by
\Be
p_\mu[m]=
\frac{N\exp (-\Delta H_m(\bm{v}^{(\mu)}[m])}
{\displaystyle\sum_\mu\exp (-\Delta H_m(\bm{v}^{(\mu)}[m])}\ ,
\Ee
where $\Delta H_m$ is a current slice of the remaining part of the Hamiltonian. 
Using this
Configuration with Probability: $\{\bm{v}^{(\mu)}[m], p_\mu[m]\}$ as the target
data, RBM parameters are tuned up ($\bm{k}_m \rightarrow \bm{k}_{m+1}$),
\Be
\left[\{\bm{v}^{(\mu)}[m], p_\mu[m]\} \Longleftrightarrow {\rm RBM}(\bm{k}_m)
\right]
\Longrightarrow{\rm RBM}(\bm{k}_{m+1})\Longrightarrow
\{\bm{v}^{(\mu)}[m+1]\}\ ,
\Ee
and the output data set $\{\bm{v}^{(\mu)}[m+1]\}$ 
by RBM $(\bm{k}_{m+1})$ is expected to obey the probability distribution,
\Be
P_{m+1}(\bm{v}) \propto \exp(-H_0(\bm{v})) \exp(-H_{m}(\bm{v})).
\Ee
Then this data set works for the
input configuration set for the next sliced step and is coupled with 
probability $p_\mu[m+1]$ defined through $\Delta H_{m+1}$.

In this serial procedures of learning, the set of configuration is
simultaneously updated. The set is updated at each step of the
steepest descent move of the machine through the contrastive
divergence iteration. At the stationary point of the machine, the
final set of configuration is used as the initial set of configuration
for the next slice, that is, each configuration is assigned the 
probability coming from the next sliced Hamiltonian effect.

Actually our slicing order respects the range of interactions
as follows.
Starting with the nearest neighbor configurations, where the probability
of configuration is all 1 (constant), we add the non-nearest neighbor
interactions of range 2, but sliced (divided) by some number. We proceed RBM
learning slice by slice, to reach the range 2 full interactions.
Then we add range 3 interactions, again with a slice. Proceeding
this way further, finally we reach the maximum range interactions,
which is 9 in this article.

Practical and full analysis of RBM learning procedures are
reported in the future full paper and here we show the
tuned RBM parameters and its evaluation by checking the susceptibility
estimates.
The size of the system is 128 spins (the number of visible variables
$\bm{v}$). The RBM links contains up to $k_{12}$, that is, 
RBM has 13 machine parameters.
The total number of configurations for input is $1024$.
We take 64 random number series to get averaged 
RBM machine parameters. The initial values of parameters
are taken to be normal values $k_0=1,k_1=1, k_n=1/n^2  \ ({\rm for\ } n>1)$. 
The total structure of the likelihood function in the multi-dimensional
space of $\bm{k}$ will not be discussed here.
In fact, 64 machines give well-converged results and we
take averaged machine parameters to define the tuned up RBM 
in the following results. 

\begin{table}
    \includegraphics[clip, width=12cm]{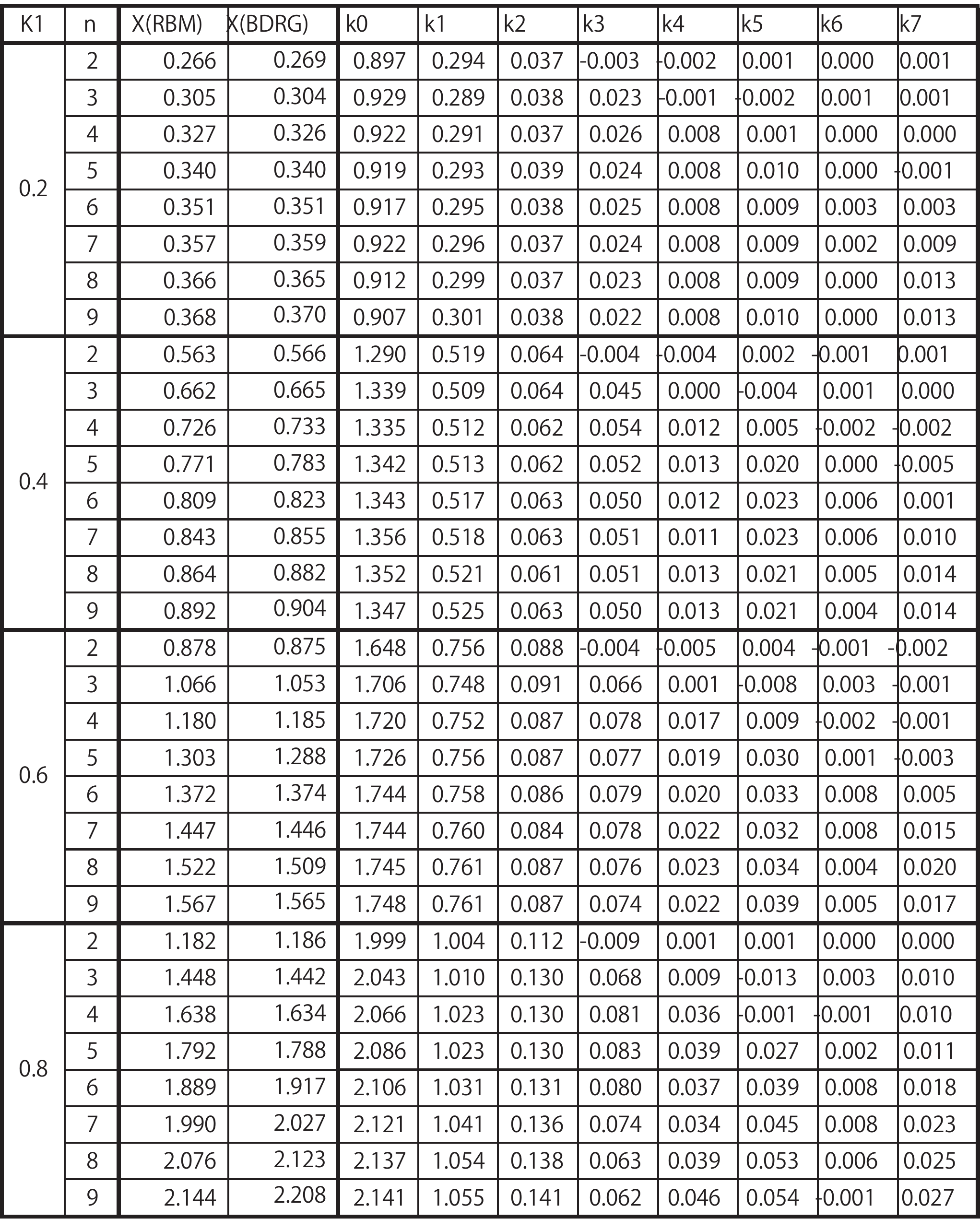}
    \caption{Tuned Restricted Boltzmann Machines and their Evaluation.}       
\end{table}

Table~1 is the results of the averaged RBM parameters, where 
$K_1$ is the nearest neighbor coupling constant and 
$n$ is the maximum rang of the target LRI model interactions.
For $n>7$, optimized $k_n$ are all small numbers and are not listed 
in the table.

In order to evaluate the quality of tuned RBM, 
we compare the susceptibility given by RBM 
with those calculated by the  
Block Decimation Renormalization Group (BDRG).
We refer to a half of the logarithm of susceptibility $\chi$,
\Be
X = \log(\chi)/2\ ,\ \ 
\chi=\frac{1}{2j_{\rm M}}\left\langle \left(\sum_i v_i\right)^2 \right\rangle.
\Ee
For the nearest neighbor case, $X$ coincides with the coupling constant,
\Be
X = K_1\ ,
\Ee
exactly in the infinite size limit.

\begin{figure}[htbp]
    \includegraphics[clip, width=10cm]{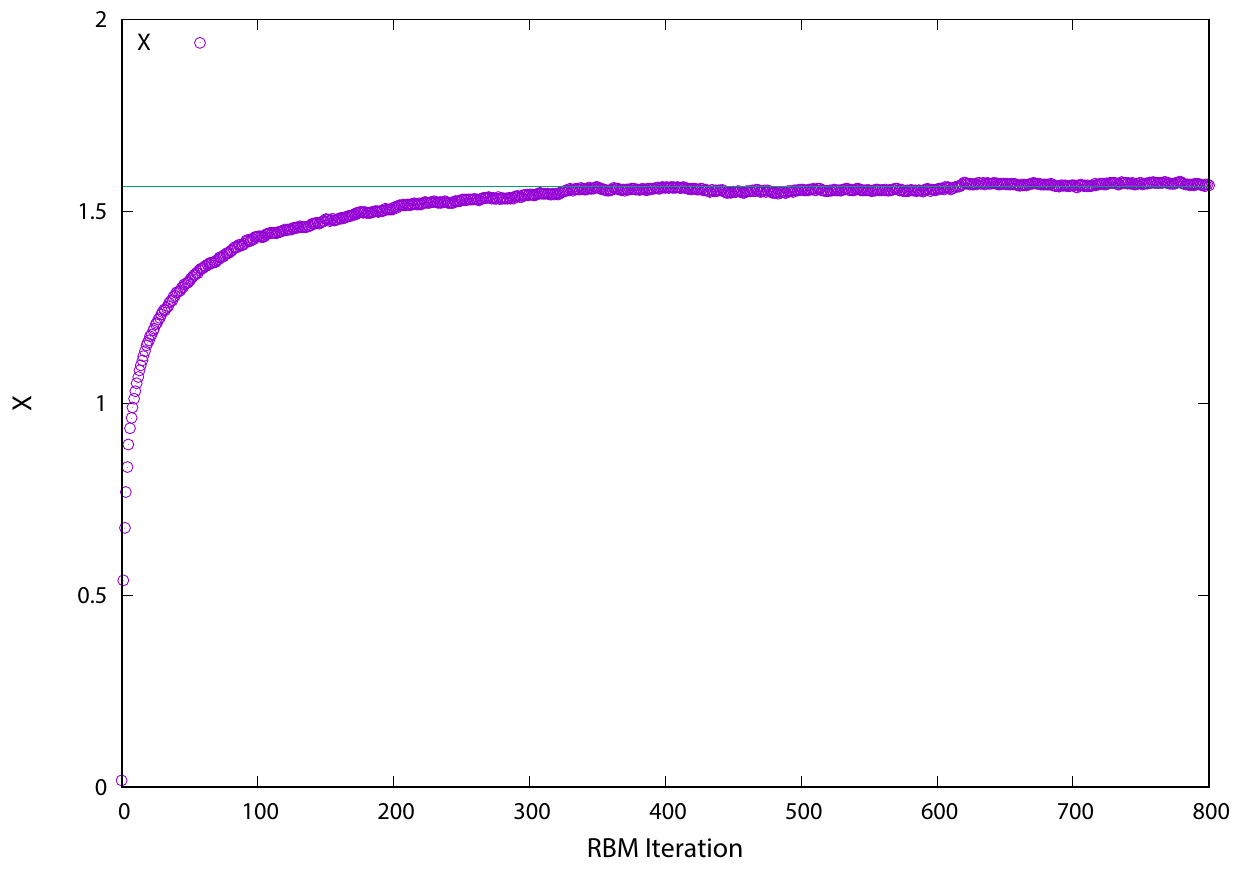}
    \caption{RBM iteration of output data.}       
    \label{fig:RMBoutput}
\end{figure}

Starting with a set of 1024 perfectly random spin configurations 
(high temperature limit ensemble), 
we operate the tuned up RBM.
We evaluate the susceptibility, step by step, which is seen in 
Fig.~3 as an example, where the target system is $K_1=0.6$ and $n=9$.
The value $X$ starts from the vanishing value of random spins
and it increases rather quickly. Finally it slowly
approaches towards the target value (1.565 in this case) which is 
drawn by a straight line.
After the equilibrium, the {\sl thermal} fluctuation is observed,
whose size will be argued in a separate paper.
After the onset of thermalized equilibrium, we read out the
parameter $X$ of the RBM by averaging over 100 iterations.

\begin{figure}[htbp]
    \includegraphics[clip, width=10cm]{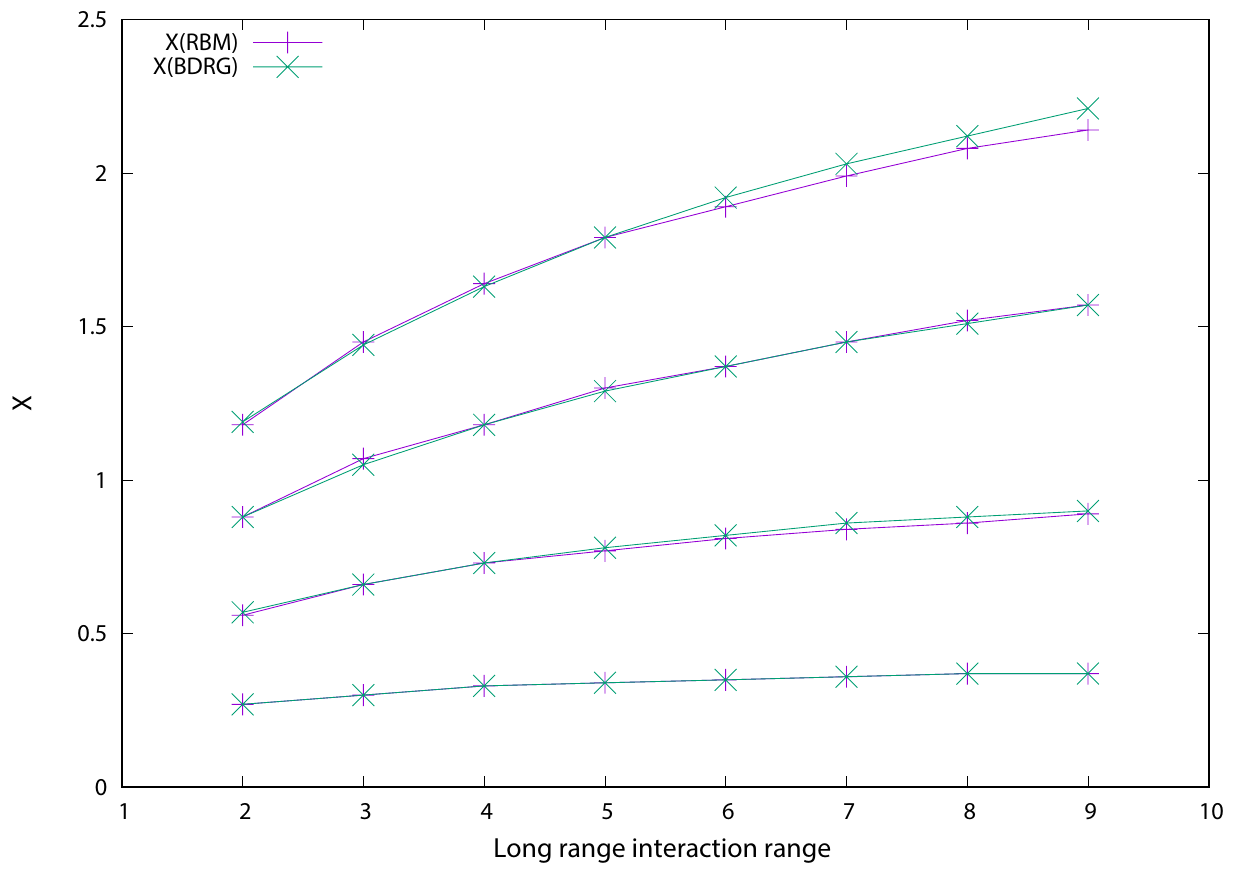}
    \caption{Evaluation of RBM by comparing with BDRG.}       
    \label{fig:BDRGcomparison}
\end{figure}

The results are listed in Table~1 and are 
shown in Fig.~4, where all data of four $K$ values
are plotted ($K=0.2, 0.4, 0.6, 0.8$ from bottom to top). 
The coincidence looks very good and our tuned RBM well
reproduce the LRI model results for the wide range of
parameter values of $K$ and $n$.
As for large susceptibility region, however, there appears small 
differences, some part of which might come from the fact
that our system is finite, periodic 128 spins, and shortage of
the number of input configurations and/or iteration sets
 of 
learning and evaluation. These will be discussed in a separate
paper.

It should be noted here that
the susceptibility is just one physical 
quantity though it is most important, and we will investigate 
the RBM output configurations in detail to further check the total
equivalence or quality of probability distribution.
Also we will clarify the intimate relation between
RBM and renormalization group method through 
multi-layer RBM systems, which will give us a new viewpoint to
understand the physical features of LRI models.
 
We thank fruitful discussions with Shin-Ichiro Kumamoto,
Hiromitsu Goto and Daisuke Sato. 
This work is first motivated by the general
lecture given by Muneki Yasuda
and we thank him much for telling us basic notions
of recent development of deep machine learning.

This work was partially supported by 
JSPS KAKENHI Grant Number 25610103
and the 2015th Research Grant of Yonago National College of Technology.

\end{document}